\documentclass[%
  reprint,
  superscriptaddress,
  aps,
  prl,
  amsmath,
  amssymb,
  preprintnumbers
]{revtex4-2}

\usepackage{graphicx}
\usepackage[x11names,dvipsnames]{xcolor}
\usepackage{ulem}
\usepackage{dcolumn}
\usepackage{bm}
\usepackage[colorlinks=true,citecolor=blue,linkcolor=blue,urlcolor=blue]{hyperref}

\begin{document}

\preprint{IPPP/26/89}

\title{Probing Ultralight Dark Matter at the Electroweak Scale}

\author{Martin Bauer and Sreemanti Chakraborti}
\affiliation{Institute for Particle Physics Phenomenology, Department of Physics, Durham University, Durham, DH1 3LE, United Kingdom}

\date{\today}

\begin{abstract}
Ultralight dark matter provides an explanation for astrophysical evidence for dark matter at all scales. This non-thermal dark matter requires a misalignment mechanism to reproduce the observed matter energy density of the Universe and therefore induce oscillating, time-dependent signals. We determine the sensitivity of current and future collider measurements of the width of the Z and Higgs bosons, and show that CP odd observables such as dipole moments in $Z\to \tau^+\tau^-$ decays provide a better sensitivity to ultralight dark matter than atomic clocks or equivalence principle tests.   
\end{abstract}

\maketitle

\section{Introduction}
Ultralight bosonic dark matter (ULDM) provides a wave-like alternative to conventional particle dark matter~\cite{Hu:2000ke,Hui:2016ltb,Antypas:2022asj}. For a light scalar field $\phi$ with an approximately quadratic potential, a non-thermal relic abundance can arise through the misalignment mechanism: once the Hubble expansion rate drops below the field mass, the initially displaced field begins to oscillate and its average energy density subsequently redshifts as non-relativistic matter~\cite{Arias:2012}. In a region smaller than the coherence length of the galactic halo, the resulting field can be approximated by $ \phi(t,\mathbf{x})= \phi_0
  \cos\!\left(m_\phi t-\mathbf{k}_\phi\!\cdot\!\mathbf{x}+\theta\right)$
where $\phi_0= \sqrt{2\rho_\phi}/m_\phi$ is set by the local dark matter energy density$\rho_\phi$. For a scalar comprising all of the dark matter, structure formation is suppressed below a mass-dependent scale. Therefore, observations of small-scale structure constrain very small masses~\cite{Hu:2000ke,Irsic:2017,Kobayashi:2017}. The large occupation numbers characteristic of ULDM allows for a classical-field treatment for masses $10^{-22}\,{\rm eV}\lesssim m_\phi\lesssim1\,{\rm eV}$, which correspond to oscillation periods $T_\phi=2\pi/m_\phi$ spanning years to a few femtoseconds.

Interactions with Standard Model (SM) particles convert the oscillating background into time-dependent observables. Linear scalar couplings to electrons or photons can produce fractional variations $\delta m_e/m_e$ and $\delta\alpha/\alpha$~\cite{Arvanitaki:2015,VanTilburg:2015,Stadnik:2016,Antypas:2022asj}.Quantum sensor experiments such as atomic spectroscopy, optical clocks and clock-cavity comparisons constrain such variations over the frequency ranges accessible  ~\cite{Arvanitaki:2015,VanTilburg:2015,Stadnik:2016,Kennedy:2020, Bauer:2024hfv}. Laser interferometers probe complementary ranges of scalar masses and couplings~\cite{Vermeulen:2021,Aiello:2022,Gottel:2024}. In addition, light scalars may mediate composition-dependent forces, allowing equivalence-principle and fifth-force tests to constrain the same underlying interactions~\cite{Hees:2018, Berge:2018, Touboul:2022}. 

Collider experiments offer a qualitatively different approach. Interactions of ULDM with a heavier state can imprint its oscillation on mediator masses~\cite{Guo:2023, Bauer:2026dgu, Fieg:2026zdr, Brade:2026ncf}, couplings, event rates or rare-decay probabilities~\cite{Dine:2024perez, Bigaran:2026}. In contrast to small structure observations and clock and interferometer bounds, these probes do not rely on resolving the periodicity of the signal, but are sensitive even to very fast oscillations, which e.g. affects the peak structure of the resonance. Here, we consider the effect of an oscillating dark matter field on the $Z-$ and Higgs boson resonance peaks and compare the results with constraints from quantum sensor measurements. We discuss several scenarios that can be considered ultraviolet (UV) completions of the effective model used to describe variations in the fine-structure constant. We point out the unique sensitivity of CP-odd observables, such as spin correlations in $Z\to \tau^+\tau^-$ decays. We show that in contrast to CP-even collider observables, CP-odd observables can be more sensitive than measurements with quantum sensors.

\section{Oscillating SM parameters} 
We consider interactions between an ultralight dark matter field and the unbroken $SU(2)_L\times U(1)_Y$ gauge fields of the SM, 
\begin{align}\label{eq:gaugevar}
    \mathcal{L}\supset c_B \frac{\phi}{\Lambda}B_{\mu\nu}B^{\mu\nu}+c_W \frac{\phi}{\Lambda}W^a_{\mu\nu}W^{a, \mu\nu}\,.
    \end{align}
In the broken phase, these interactions induce a time-dependent fine-structure constant 
\begin{align}
    \frac{\delta \alpha}{\alpha}&=4\left(c_Bc_\theta^2+c_W \left(s_\theta^2-\frac{7\alpha}{4\pi}\right)\right)\frac{\phi}{\Lambda}\,,
\end{align}
where the last term is the one-loop correction induced by the charged $W-$boson loop and $s_\theta =\sin(\theta)$ is the Weinberg angle. In addition, electroweak masses and mixing angles can oscillate, too
\begin{align}\label{eq:EWvar}
 \frac{\delta M_W}{M_W}&=2c_W\frac{\phi}{\Lambda}\,,\quad 
     \frac{\delta M_Z}{M_Z}=2\left(c_Wc_\theta^2+c_Bs_\theta^2\right)\frac{\phi}{\Lambda}\,,\notag\\
 \frac{\delta s_\theta}{s_\theta}&=4c_\theta^2\left(c_B-c_W\right)\frac{\phi}{\Lambda}\,.
\end{align}
Even for vanishing $\delta \alpha=0$, the masses of the electroweak gauge bosons and the Weinberg angle can oscillate. 
We further consider couplings of dark matter to the Higgs sector
\begin{align}
\mathcal{L}\supset c_H \mu^2\frac{\phi}{\Lambda}H^\dagger H-c_\lambda \lambda \frac{\phi}{\Lambda} (H^\dagger H)^2\,.
\end{align}
These interactions can lead to a time-dependent vacuum expectation value, inducing oscillating weak gauge boson masses, fermion masses and an oscillating Fermi constant, 
\begin{align}\label{eq:hvar}
\frac{\delta M_Z}{M_Z}=\frac{\delta M_W}{M_W}=\frac{\delta m_f}{m_f}=-2\frac{\delta G_F}{G_F}&=\frac12(c_H-c_\lambda)\frac{\phi}{\Lambda}\,.
\end{align}
The time-dependence of the fine-structure constant is induced by loops of charged fermions and $W$ bosons, so that 
\begin{align}
\frac{\delta \alpha}{\alpha}=\frac{11}{12\pi}\alpha (c_H-c_\lambda)\frac{\phi}{\Lambda}\,.
\end{align}
In the limit $c_H=c_\lambda$, the Higgs vev is time-independent and  only the Higgs mass oscillates,
\begin{align}
\frac{\delta M_h}{M_h}&=c_H\frac{\phi}{\Lambda}\,.
\end{align} 
In this case the fine-structure constant would only become sensitive at two-loop order with Higgs corrections to $W-$boson and top loops, 
\begin{align}
\frac{\delta \alpha}{\alpha}\approx -\frac{\alpha^2}{64\pi^2s_\theta^2}\left(\frac{M_h^2}{M_W^2}+2N_C\frac{m_t^2}{M_W^2}\right)c_H\frac{\phi}{\Lambda}
\end{align}

\begin{figure}
\includegraphics[width=.3\textwidth]{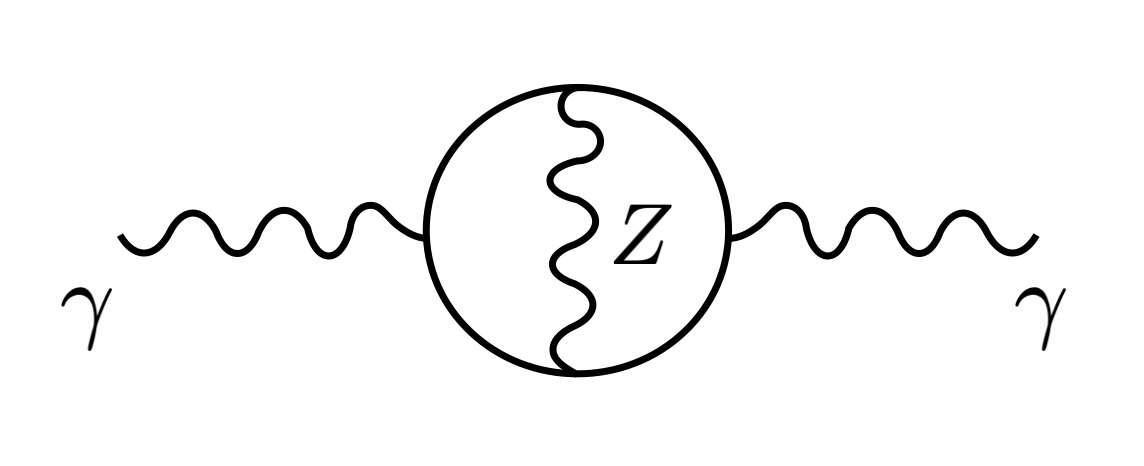}
\caption{\label{fig:Zdiagram}Example for two-loop diagrams contributing to the beta function for $\alpha$ with an internal Z propagator.}
\end{figure}
Two loop corrections to the fine-structure constant are unavoidable if the Z mass or couplings change, because the Z propagator enters the beta function for $\alpha$ at two loops, see Figure \ref{fig:Zdiagram}. Even if dark matter interactions only arise through operators with additional derivatives, such that the leading contribution to the fine-structure constant is momentum suppressed $\delta \alpha \propto q^2$, contributions from integrating out the Z-boson at $q^2=M_Z^2$ induce a correction without this suppression. If the dark matter only affects the Z vertex instead, it enters quadratically in the beta function for $\alpha$, but linearly in the Z decay amplitude. However, the partial width of the Z boson affected by this coupling has a quadratic dependence again.  
\begin{align}
   \frac{ \delta \alpha}{\alpha} \approx \frac{\alpha^2}{(16\pi^2)^2}\frac{\delta M_Z}{M_Z}.
\end{align}
If instead a CP-odd operator couples to an ultralight dark matter field, the dependence of CP-odd observables can be linear in the new physics coupling, whereas a variation of the fine-structure constant depends quadratically on this coupling. An example of this type of operators are lepton dipoles
\begin{align}
\mathcal{L}&\supset c_{DB}\frac{i\phi}{\Lambda^3}(\bar L\sigma^{\mu\nu}\ell_R)H B_{\mu\nu}\notag\\
&\quad+c_{DW}\frac{i\phi}{\Lambda^3}(\bar L\sigma^{\mu\nu}\tau^a\ell_R)H W^a_{\mu\nu}+h.c.
\end{align}
These operators induce corrections to the photon and Z-lepton vertices
\begin{align}
\Gamma_\ell^{\gamma,\nu}&=ie\Big[
Q_\ell\gamma^\nu
+
\frac{\widetilde d_\ell^{\,\gamma}}{2m_\ell}
\gamma_5\sigma^{\nu\rho}q_\rho
\Big]\\
\Gamma_\ell^{Z,\nu}&=ie\Big[
v_\ell\gamma^\nu-a_\ell\gamma^\nu\gamma_5
+
\frac{\widetilde d_\ell^{\,Z}}{2m_\ell}
\gamma_5\sigma^{\nu\rho}q_\rho
\Big]\,,
\end{align}
where 
\begin{align}
\widetilde d_{\ell}^{\,\gamma}
&=
-\frac{2\sqrt{2}}{e}\,
\frac{m_\ell v\,\phi}{\Lambda^3}
\left(
c_\theta c_{DB}-s_\theta c_{DW}
\right),
\\
\widetilde d_{\ell}^{\,Z}
&=
\frac{2\sqrt{2}\,}{e}\,
\frac{ m_\ell v\,\phi}{\Lambda^3}
\left(
s_\theta c_{DB}+c_\theta c_{DW}
\right).
\end{align}
For $Z\to\tau^+\tau^-$ decays, these dipoles enter linearly in CP-odd correlations between the tau decay-products~\cite{Bernreuther:1991xe}. The contribution to $\delta \alpha$ depends quadratically on the dipole

\begin{align}\label{eq:CPoddalpha}
\frac{\delta\alpha}{\alpha}
&=
\frac{\alpha}{2\pi}
\ln\!\left(\frac{M_Z}{m_\ell}\right)
(\widetilde d_\ell^\gamma)^2
\left(1+\delta_2^{\rm QED}\right)
\notag\\
&+
\frac{\alpha^2}{64\pi^2}
\frac{M_Z^2}{m_\ell^2}
\Big[
 F_{\gamma\gamma}^{\rm EW}
(\widetilde d_\ell^\gamma)^2
+ 2 F_{\gamma Z}^{\rm EW}\,
\widetilde d_\ell^\gamma\widetilde d_\ell^Z
+ F_{ZZ}^{\rm EW}
(\widetilde  d_\ell^Z)^2
\Big]\notag\\
&+
\frac{\delta\alpha_{\rm local}}{\alpha}
+\cdots\,. 
\end{align}
The leading logarithmic contribution is the one-loop double photon-dipole insertion~\cite{Jenkins:2017dyc}. The coefficient $\delta_2^\text{QEC}$ accounts for photons dressing of the corresponding one-loop diagram and therefore contains an additional loop factor and additional logarithms.
Electroweak matching coefficients $F_{\gamma\gamma}^{\rm EW}, F_{\gamma Z}^{\rm EW}, F_{ZZ}^{\rm EW}$ depend on the two-loop diagrams. The scale dependence of these loop functions cancel with contributions from higher order local operators. We consider them to be order one here by dimensional analysis,
but their finite values require an explicit two-loop electroweak matching calculation. Nuclear dipole operators have been proposed as potential probes for axions ~\cite{Graham:2013gfa}.

\section{Collider bounds on ultralight dark matter}

\paragraph{Z width and pole mass.}
Even for high oscillation frequency, an oscillating DM background can leave observable imprints in the line shape of resonances. The line shape of the $Z$-resonance is sensitive to ULDM interacting with electroweak gauge bosons~\eqref{eq:gaugevar} as well as with the Higgs potential~\eqref{eq:EWvar}. Since the LEP measurements constrain both the pole mass and the decay width of the $Z$ boson with high precision, the experimentally allowed oscillation amplitude is expected to satisfy $\delta M_Z\ll \Gamma_Z\simeq 2.5\,{\rm GeV}$. In this regime, the oscillation does not produce a resolved double-peak structure around the central resonance~\cite{Guo:2022vxr,Bauer:2026dgu}, but instead manifests itself as a distortion of the phase-averaged resonance profile. The sensitivity of the analysis is therefore entirely determined by our ability to accurately model the shape of the measured cross section in the vicinity of the $Z$ pole.

To determine the sensitivity to oscillating dark matter, we fit LEP data~\cite{L3:1993zvy,DELPHI:1994bya} using the resonant Z-production cross section convoluted with the dominant corrections due to initial-state radiation,
\begin{align}
\sigma_Z(s)
=
\int_{x_{\rm min}}^{1}
dx\,
\rho(x)\,
\sigma_0(xs)\,,
\end{align}
where $x=s'/s$ denotes the fraction of the original centre-of-mass energy remaining after photon emission and $\rho(x)$ is the standard QED radiator function~\cite{KuraevFadin1985}
\begin{align}
\rho(s,x)
=
\beta(s)(1-x)^{\beta(s)-1}
\delta^{\rm sv}(s)
+
\delta^{\rm h}(s,x)
\end{align}
with 
\begin{align}
\beta(s)&=-\frac{2\alpha}{\pi}\left[{\rm ln}\,x_e+1\right]\,,\\
\delta^{\rm sv}(s)&=1+\frac{\alpha}{2\pi}\left[-3\,{\rm ln}\,x_e+2\pi^2/3-4\right]\,,\\
\delta^{\rm h}(s,x)&=\frac{\alpha}{\pi}(1+x)\left[{\rm ln}\,x_e+1\right]\,,
\end{align}
and $x_e=m_e^2/s$.
For the cross section we use 
\begin{align}\label{eq:S0}
\sigma_0(s) = \frac{4\pi\alpha^2}{3}
\left(
\frac{r_T^{\gamma f}}{s}
+
\frac{s\,r_T^f+(s-M_Z^2)j_T^f}
{(s-M_Z^2)^2+M_Z^2\Gamma_Z^2}
\right)\,,
\end{align}
 which includes corrections from $\gamma-Z$ interference off resonance with the SM prediction $j_T^\text{had}=0.22$ and $r_T^\text{had}=2.97$ for hadronic final states. Effects from the LEP beam-energy spread only affect the uncertainty of the $Z$-boson width at the percent level and can be neglected. The mass of the $Z$-boson enters the cross section \eqref{eq:S0} and the total width of the $Z$-boson $\Gamma_Z$. The effect of oscillation due to interactions with the light dark matter field can be captured by replacing $M_Z\to M_Z(1+A_1^Z\cos(\theta))$ and using \eqref{eq:EWvar} or \eqref{eq:hvar}~\footnote{For a dark matter field that interacts quadratically with SM fields, the corresponding time-dependence of the Z-mass is $M_Z\to M_Z(1+A_2^Z\big(1+\cos(2\theta))\big)=M_Z(1+A_2^Z\cos(2\theta)\big)$ after absorbing the constant into $M_Z$. The bounds on $A_2^Z$ and $A_2^H$ are therefore the same as $A_1^Z$ and $A_1^H$ throughout the paper}. The phase-averaged cross section then reads
\begin{align}
\overline{\sigma}(s)=\frac{1}{2\pi}
\int_0^{2\pi}
d\theta\,
\sigma_Z(s)\big\vert_{M_Z\to M_Z\big(1+A_1^Z\cos(\theta)\big)}\,.
\end{align}
Larger values of the amplitudes $A_1^Z\propto 1/(\Lambda \,m_\phi)$ and $A_2^Z\propto 1/(\Lambda \,m_\phi)^2$ lead to increasingly visible
distortions of the predicted line shape and consequently a deterioration
of the goodness-of-fit to the LEP data. The oscillation amplitude is therefore constrained by requiring the resulting line shape to remain compatible with the LEP measurements within their experimental uncertainties. For each value of $A_1^Z$, the corresponding phase-averaged line shape is compared with the LEP data using a $\chi^2$
 statistic. The 95\% confidence-level upper limit on the amplitude is then defined by $\Delta\chi^2=\chi^2(A^Z_i)-\chi^2(A^Z_i=0)=3.84$. The result of the fit to the Z-boson line width can then be expressed as a bound on the amplitude 
 $A_1\leq 1.41\cdot 10^{-3}\,$. For comparison, atomic clock bounds on $\alpha$ variation imply equivalent sensitivities of $A_1\lesssim 10^{-17}$~\cite{Antypas:2022asj}.

A future measurement of the Z-boson lineshape at the FCC-ee can significantly improve this bound~\cite{FCC:2025uan}. We estimate the projected sensitivity based on the proposed three-point $Z$-pole scan with centre-of-mass energies and luminosities $(\sqrt{s}/\text{GeV},\mathcal{L}/{\rm ab}^{-1} )= (87.9\,,40), (91.2\,,125),(94.3, 40)$. , and generate pseudo-data from the ISR-convoluted S-matrix line shape with the statistical uncertainty given by $\delta\sigma_i^{\rm stat}=\sqrt{\sigma_i/\mathcal{L}_i}$. 
While the expected statistical uncertainty of the Z-pole mass measurement at FCC-ee is a few keV, the systematic uncertainty is currently expected to be 100 keV. We define a pessimistic scenario including the systematic uncertainty and an optimistic scenario in  which the statistical uncertainty dominates and find $A_1\leq 1.02\cdot 10^{-4}$ (systematically dominated) and $A_1\leq 1.98\cdot 10^{-5}$ (statistically dominated), respectively.

\paragraph{Higgs width and pole mass.} A future FCC-ee measurement of the width of the Higgs boson would provide additional information. In contrast to the $Z$-line shape measurement, which determines $M_Z$ and $\Gamma_Z$ from a resonance scan of the energy-dependent cross-section around the $Z$ pole, the FCC-ee Higgs mass measurement relies on the recoil mass technique in the Higgsstrahlung process $e^+e^-\to ZH$ at a fixed centre-of-mass energy of $\sqrt{s}=240~\mathrm{GeV}$~\cite{Li:2025kjy}. The Higgs mass is reconstructed from the kinematics of the associated $Z\to\mu^+\mu^-$ decay via the recoil mass, independent of the Higgs decay mode. As a result, the observable is not a Breit--Wigner resonance but a detector-limited recoil mass peak, whose width ($\sim0.5~\mathrm{GeV}$) is much larger than the intrinsic Higgs width ($\Gamma_H\approx4~\mathrm{MeV}$). In principle this process is sensitive to both an oscillating Higgs mass and an oscillating Z-boson mass, but the latter enters only through the smooth Higgsstrahlung production cross section and has a negligible impact on the recoil mass shape. The consequence of an oscillating Higgs mass would produce a time-dependent displacement of the recoil mass peak, which appears experimentally as an effective broadening after averaging over many oscillation periods. In the following, we consider the Higgsstrahlung channel with $Z\to\mu^+\mu^-$, adopting the FCC-ee baseline assumptions of $\sqrt{s}=240~\mathrm{GeV}$, an integrated luminosity of $10.8~\mathrm{ab}^{-1}$, $\sigma_{ZH}=200~\mathrm{fb}$, $\mathrm{BR}(Z\to\mu^+\mu^-)=3.366\%$, and a muon selection efficiency of $74.66\%$, which corresponds to an expected signal yield of approximately $5.4\times10^4$ events. The signal recoil mass distribution is extracted by digitising the high-BDT signal-plus-background and background templates from Ref.~\cite{Li:2025kjy}, subtracting the background, and fitting the resulting signal with a Gaussian of width $\sigma_{\rm eff}\simeq0.51~\mathrm{GeV}$, consistent with an independent estimate from the inclusive recoil-mass spectrum. The background is modelled by an asymmetric two-sided Gaussian, yielding $S\approx1.7\times10^4$, $B\approx3.3\times10^3$ over the fitted mass window and a statistical sensitivity consistent with the projected FCC-ee precision.
An oscillating Higgs mass is incorporated by replacing $M_H\rightarrow M_H(1+ A_1^H\cos(\theta)\big)$ and averaging over the oscillation phase in the binned recoil-mass spectrum, results in a constraint $A_1^H\leq 9.5\times 10^{-4}$. For comparison, the equivalent bound inferred from atomic clocks is $A_1^H\lesssim  10^{-11}$~\cite{Antypas:2022asj}. 



We further estimate the sensitivity of a direct Higgs resonance scan in $\mu^+\mu^-\to h\to X$ at a future muon collider. Following Ref.~\cite{Han:2012rb}, we model the resonance using a Breit-Wigner distribution convolved with a Gaussian beam-energy spread. We adopt their benchmark with a beam energy spread resolution of $R=0.01\%$ and a projected Luminosity of $\mathcal{L}=0.5\ {\rm {fb}^{-1}}$, and we perform a 20-point scan over $m_H\pm 30\ {\rm MeV}$, including the quoted $b\bar{b}$ and $WW^*$ signal and background rates. The predicted event rates are averaged over the oscillation phase, and the $\chi^2$ analysis yields the a 95\% C.L of $\delta M_H^{95}\simeq 5.79~{\rm MeV}$, resulting in a projected sensitivity to an amplitude $A^{H}\le 4.6\times 10^{-5}$.

\begin{figure}[t]
  \centering
  \includegraphics[width=0.99\columnwidth]{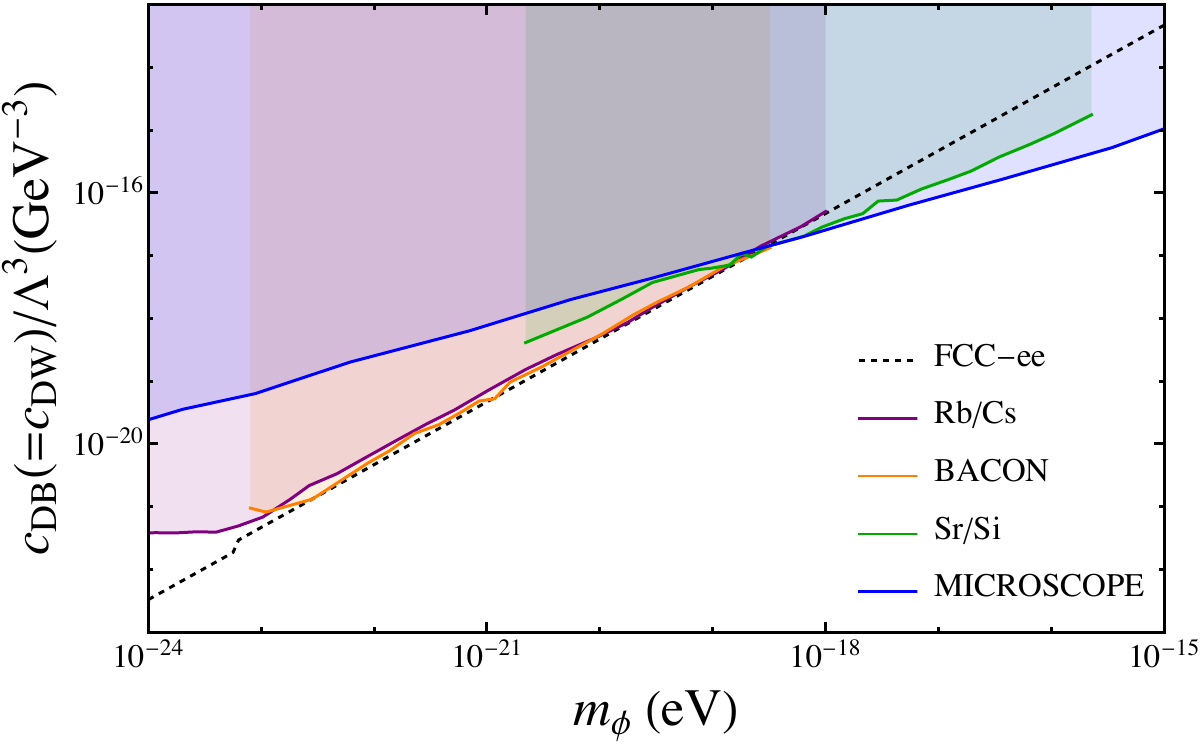}
  \caption{Projected FCC-ee sensitivity (black dashed) to the tau dipole operators along $c_{DB}=c_{DW}$, compared with constraints from atomic clock and optical cavity bounds (Rb/Cs, BACON, and Sr/Si) and MICROSCOPE. Shaded regions are excluded by existing measurements.}
  \label{fig:QSvsCPodd}
\end{figure}

\begin{figure}[t]
  \centering
  \includegraphics[width=.9\columnwidth]{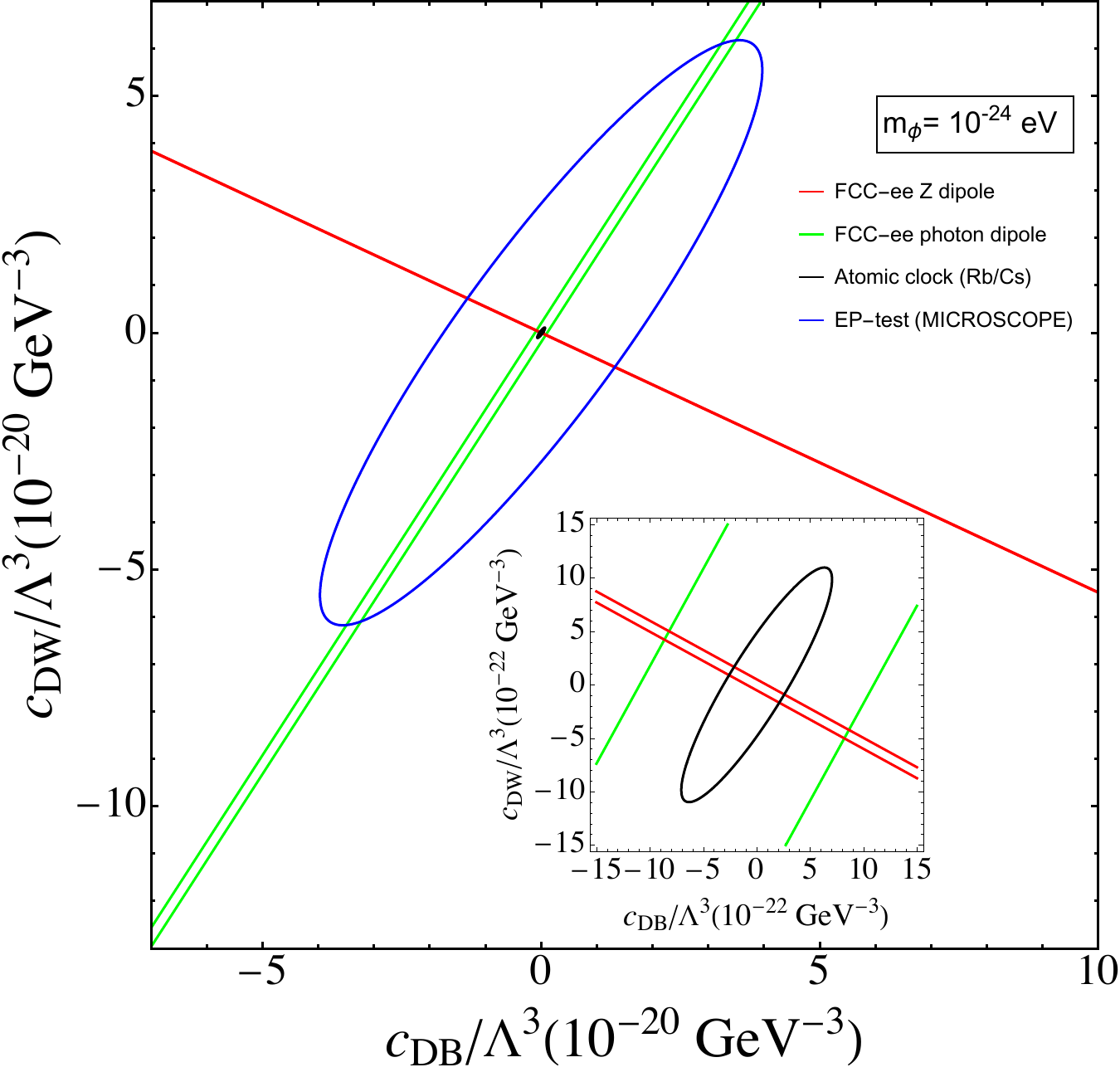}
  \caption{Constraints on $c_{DB}/\Lambda^3$ and $c_{DW}/\Lambda^3$. Green and red lines denote the projected FCC-ee photon and $Z$-dipole sensitivities, respectively, while black and blue ellipses show the atomic clock and MICROSCOPE bounds. The inset magnifies the region around the clock constraint.}
  \label{fig:main}
\end{figure}

\paragraph{CP odd observables.}
Whereas CP-even collider observables such as the width and mass of the Higgs and the Z-boson are generically less sensitive to ULDM couplings than quantum sensor probes for an oscillating fine-structure constant, the situation is different if dark matter interacts with CP-odd operators. We consider searches for a photon dipole by Belle~\cite{Belle:2021ybo}, which constrains $-3.33\times10^{-3}
< \operatorname{Re}\widetilde d_\tau^{\,\gamma}
< 1.10\times10^{-3}$, and searches for the Z-dipole in $Z\to \tau^+\tau^-$ decays performed by ALEPH~\cite{ALEPH:2002kbp}, with $|\text{Re}\widetilde d_\tau^{\,Z}|
< 0.91\times10^{-3}$. 
\begin{figure}[t]
  \centering
  \includegraphics[width=.95\columnwidth]{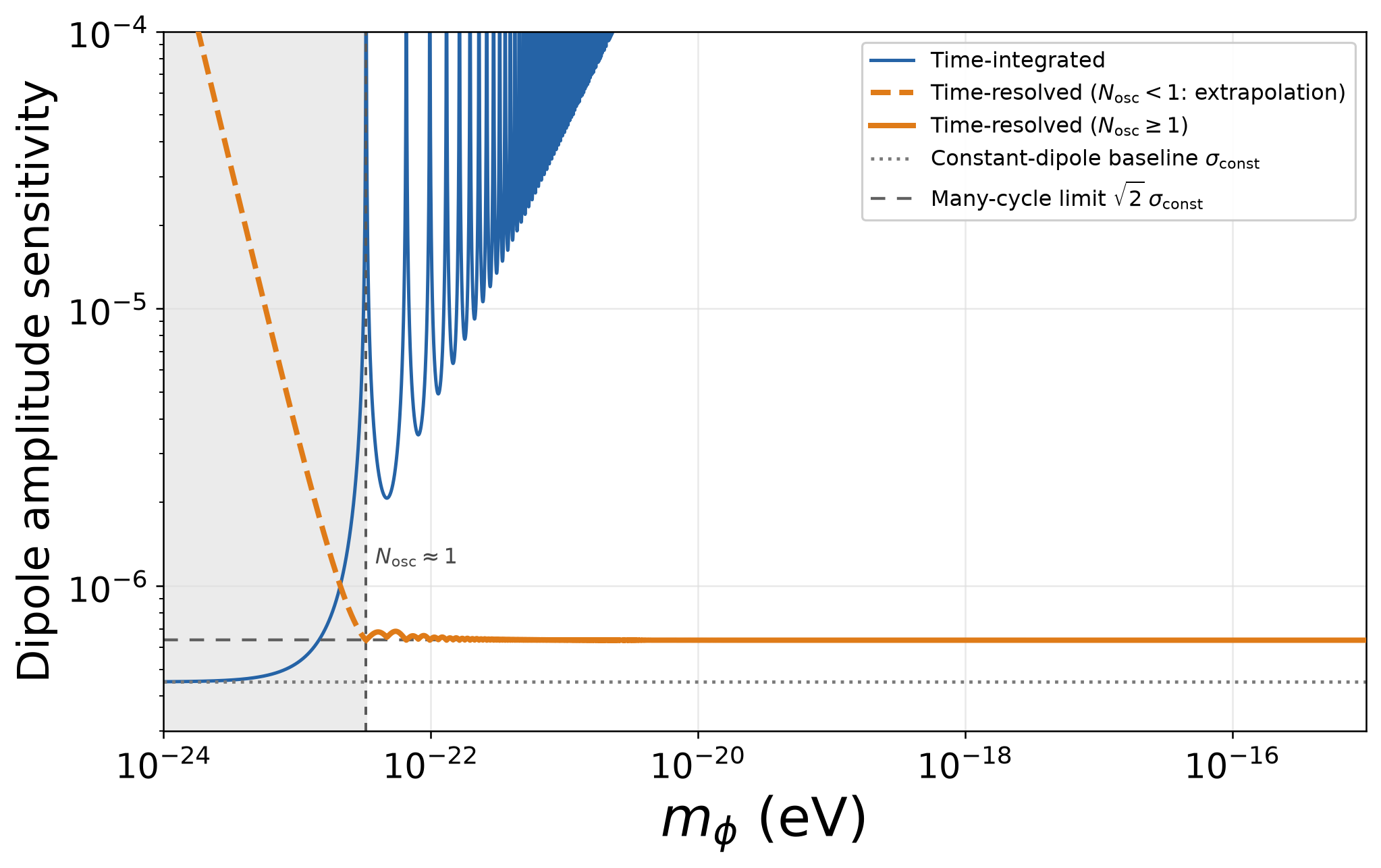}
  \caption{Projected FCC-ee sensitivity to an oscillating CP-odd $\tau$ $Z$-dipole as a function of the ULDM mass $m_\phi$, assuming a 4-year observation period. The blue curve shows the phase-optimised sensitivity of a time-integrated analysis, including the suppression from averaging over multiple oscillations. The orange curve shows the estimated time-resolved sensitivity obtained from a sinusoidal quadrature fit with a freely varying constant offset. Its dashed portion corresponds to $N_{\rm osc}<1$, where the oscillation is not fully resolved. The grey dotted line denotes the projected constant-dipole sensitivity, $\sigma_{\rm const}$. For $N_{\rm osc}\gg1$, the time-resolved sensitivity approaches $\sqrt{2}\,\sigma_{\rm const}$, which is shown by the grey dashed line.}
  \label{fig:timeaveragedvsresolved}
\end{figure}
We further use future projections for spin correlations in $Z\to \tau^+\tau^-$ from FCC-ee~\cite{Du:2026ynu}, which are expected to lead to a significant improvement over current measurements, with a projected sensitivity of $\widetilde d_\tau^\gamma =9.6\times10^{-6}$ and $\widetilde d_\tau^Z=4.5\times10^{-7}$ for constant dipole couplings. For very light dark matter, the dipole is approximately constant over the whole observation period since 
$T_\phi \approx 1.31\, (10^{-22}~\text{eV}/m_\phi)$ years. For higher dark matter masses, the dipole induced by the ULDM oscillation takes the form $\widetilde d_\tau^Z(t)=A_Z\cos(m_\phi t+\theta)$, where the unknown phase $\theta$ affects the time-integrated signal. Averaging over an observation period $T\gg T_\phi$ suppresses the linear CP-odd observable by $|{\rm sinc}(m_\phi T/2)|$, leading to blind spots when an integer number of oscillations is completed. In contrast, a time-resolved analysis retains the oscillatory information by fitting independent sine and cosine quadratures with a floating constant offset~\cite{Zechmeister:2009js,Pomme:2020ipk}.
The corresponding sensitivity is obtained from the least-constrained direction of the quadrature covariance matrix and approaches $\sqrt{2}\sigma(\widetilde d_\tau^Z)$ in the many-cycle limit, where $\sigma(\widetilde d_\tau^Z)$ denotes the projected uncertainty on a constant $Z$-dipole coupling. Fig.~\ref{fig:timeaveragedvsresolved} illustrates the resulting dipole-amplitude sensitivity as a function of $m_\phi$, assuming a four-year FCC-ee observation baseline~\cite{FCC:2025lpp} with uniform  time-weighting. The time-integrated sensitivity (blue) exhibits oscillatory cancellations, whereas the time-resolved sensitivity (orange) remains approximately constant once at least one oscillation is observed, $N_{\rm osc}=m_\phi T/(2\pi)\gtrsim1$. The dashed orange continuation below this threshold indicates the regime where the oscillation becomes increasingly degenerate with a constant offset. The grey dotted line denotes the projected constant-dipole sensitivity quoted above. At lower ULDM masses, microwave and optical clock-comparison tests~\cite{BACON:2020ubh,Hees:2016gop,Kennedy:2020bac} impose particularly stringent constraints on the quadratic combinations of electroweak Wilson coefficients, while at higher masses the weakening of these bounds allows EP tests from MICROSCOPE~\cite{Touboul:2022} to become increasingly relevant. We show the comparison between the projected FCC-ee sensitivity with current quantum sensor bounds in Fig.~\ref{fig:QSvsCPodd}. For masses $m_\phi\lesssim 10^{-18}$ eV, FCC-ee measurements can be more sensitive than even the most precise low-energy probes available, in particular for very light masses, for which the dipole is approximately constant. We compare the allowed parameter space in Fig.~\ref{fig:main} assuming $m_\phi=10^{-24}$ eV and Wilson coefficients $c_{DB}/\Lambda^3$ and $c_{DW}/\Lambda^3$. The photon and $Z$-dipole sensitivities constrain orthogonal linear combinations of the Wilson coefficients, giving rise to the narrow bands shown in green and red. The atomic clock and the EP-violation test (MICROSCOPE) constraints instead arise from the quadratic correction to the fine-structure constant in \eqref{eq:CPoddalpha}, yielding the elliptical allowed regions in black and blue, respectively. For this comparison, we set the electroweak matching functions $F_{\gamma\gamma}^{\rm EW}=F_{\gamma Z}^{\rm EW}=F_{ZZ}^{\rm EW}=1$ and neglect the remaining contributions to $\delta\alpha/\alpha$ by setting $\delta_2^{\rm QED}=\delta \alpha_{\rm local}/\alpha=0$, therefore isolating the dipole-induced terms.

\section{Conclusion}
We present an analysis of collider observables that are sensitive to ultralight dark matter with couplings to electroweak gauge bosons, the Higgs boson or the dipole moment of tau leptons. In the broken electroweak basis, these interactions can lead to oscillations of the fine-structure constant, the Weinberg angle and masses of fermions, gauge bosons and the Higgs boson. In contrast to quantum sensor searches for variations of the fine-structure constant, the effect of ULDM oscillations on collider observables such as the line shape of the Higgs- or Z-boson can be detected even for very high oscillation frequencies because the presence of an oscillating scalar background distorts the line-shape of the Z-boson or the Higgs boson, even if the oscillation frequency cannot be resolved. Nevertheless, we find that constraints from the width of the Z-boson or current and future measurements of the width of the Higgs boson are already ruled out by atomic clock constraints or bounds from violations of the equivalence principle for the entire ultralight dark matter mass range. We argue, that this is likely to be the case for all CP-even observables at colliders, because the relevant operators inevitably induce an oscillating fine-structure constant at two-loop order, but scaling with the same coupling as the collider observables. In contrast, operators that lead to CP-odd observables, such as lepton electric dipole moments, induce a fine-structure constant at the level of the corresponding coupling squared. CP odd collider observables are therefore uniquely powerful in searching for ultralight dark matter. We demonstrate this at the example of spin-correlation measurements in $Z\to \tau^+\tau^-$ decays. For dark matter masses of $m_\phi=10^{-24}-10^{-18}$ eV, the corresponding projected FCC-ee measurements are expected to achieve a better sensitivity than atomic clocks. 

\begin{acknowledgments}
We acknowledge support from the UKRI Future Leader Fellowship DarkMAP (Ref. no. MR/Y034112/1).
\end{acknowledgments}

\bibliographystyle{apsrev4-2}
\bibliography{biblio}

@article{FCC:2025uan,
    author = "Benedikt, M. and others",
    collaboration = "FCC",
    title = "{Future Circular Collider Feasibility Study Report: Volume 2, Accelerators, Technical Infrastructure and Safety}",
    eprint = "2505.00274",
    archivePrefix = "arXiv",
    primaryClass = "physics.acc-ph",
    reportNumber = "CERN-FCC-ACC-2025-0004",
    doi = "10.1140/epjs/s11734-025-01967-4",
    journal = "Eur. Phys. J. ST",
    volume = "234",
    number = "19",
    pages = "5713--6197",
    year = "2025"
}

@article{Vermeulen:2021,
    author = "Vermeulen, Sander M. and others",
    title = "{Direct limits for scalar field dark matter from a gravitational-wave detector}",
    eprint = "2103.03783",
    archivePrefix = "arXiv",
    journal = "Nature",
    volume = "600",
    pages = "424--428",
    doi = "10.1038/s41586-021-04031-y",
    year = "2021"
}

@article{Aiello:2022,
    author = "Aiello, Lorenzo and Richardson, Jonathan W. and Vermeulen, Sander M. and Grote, Hartmut and Hogan, Craig and Kwon, Ohkyung and Stoughton, Chris",
    title = "{Constraints on Scalar Field Dark Matter from Colocated Michelson Interferometers}",
    journal = "Phys. Rev. Lett.",
    volume = "128",
    pages = "121101",
    doi = "10.1103/PhysRevLett.128.121101",
    year = "2022"
}

@article{Gottel:2024,
    author = "G{\"o}ttel, Alexandre S. and Ejlli, Aldo and Karan, Kanioar and Vermeulen, Sander M. and Aiello, Lorenzo and Raymond, Vivien and Grote, Hartmut",
    title = "{Searching for Scalar Field Dark Matter with LIGO}",
    eprint = "2401.18076",
    archivePrefix = "arXiv",
    journal = "Phys. Rev. Lett.",
    volume = "133",
    pages = "101001",
    doi = "10.1103/PhysRevLett.133.101001",
    year = "2024"
}

@article{Arvanitaki:2015,
    author = "Arvanitaki, Asimina and Huang, Junwu and Van Tilburg, Ken",
    title = "{Searching for dilaton dark matter with atomic clocks}",
    eprint = "1405.2925",
    archivePrefix = "arXiv",
    primaryClass = "hep-ph",
    journal = "Phys. Rev. D",
    volume = "91",
    pages = "015015",
    doi = "10.1103/PhysRevD.91.015015",
    year = "2015"
}

@article{VanTilburg:2015,
    author = "Van Tilburg, Ken and Leefer, Nathan and Bougas, Lykourgos and Budker, Dmitry",
    title = "{Search for Ultralight Scalar Dark Matter with Atomic Spectroscopy}",
    eprint = "1503.06886",
    archivePrefix = "arXiv",
    primaryClass = "physics.atom-ph",
    journal = "Phys. Rev. Lett.",
    volume = "115",
    pages = "011802",
    doi = "10.1103/PhysRevLett.115.011802",
    year = "2015"
}

@article{Stadnik:2016,
    author = "Stadnik, Y. V. and Flambaum, V. V.",
    title = "{Improved limits on interactions of low-mass spin-0 dark matter from atomic clock spectroscopy}",
    eprint = "1605.04028",
    archivePrefix = "arXiv",
    journal = "Phys. Rev. A",
    volume = "94",
    pages = "022111",
    doi = "10.1103/PhysRevA.94.022111",
    year = "2016"
}

@article{Kennedy:2020,
    author = "Kennedy, Colin J. and Oelker, Eric and Robinson, John M. and Bothwell, Tobias and Kedar, Dhruv and Milner, William R. and Marti, G. Edward and Derevianko, Andrei and Ye, Jun",
    title = "{Precision Metrology Meets Cosmology: Improved Constraints on Ultralight Dark Matter from Atom-Cavity Frequency Comparisons}",
    journal = "Phys. Rev. Lett.",
    volume = "125",
    pages = "201302",
    doi = "10.1103/PhysRevLett.125.201302",
    year = "2020"
}

@article{Hu:2000ke,
    author = "Hu, Wayne and Barkana, Rennan and Gruzinov, Andrei",
    title = "{Fuzzy Cold Dark Matter: The Wave Properties of Ultralight Particles}",
    eprint = "astro-ph/0003365",
    archivePrefix = "arXiv",
    journal = "Phys. Rev. Lett.",
    volume = "85",
    pages = "1158--1161",
    doi = "10.1103/PhysRevLett.85.1158",
    year = "2000"
}

@article{Hui:2016ltb,
    author = "Hui, Lam and Ostriker, Jeremiah P. and Tremaine, Scott and Witten, Edward",
    title = "{Ultralight scalars as cosmological dark matter}",
    eprint = "1610.08297",
    archivePrefix = "arXiv",
    primaryClass = "astro-ph.CO",
    journal = "Phys. Rev. D",
    volume = "95",
    pages = "043541",
    doi = "10.1103/PhysRevD.95.043541",
    year = "2017"
}

@misc{Antypas:2022asj,
    author = "Antypas, D. and others",
    title = "{New Horizons: Scalar and Vector Ultralight Dark Matter}",
    eprint = "2203.14915",
    archivePrefix = "arXiv",
    primaryClass = "hep-ex",
    reportNumber = "FERMILAB-PUB-22-262-AD-PPD-T",
    month = "3",
    year = "2022"
}

@article{Arias:2012,
    author = "Arias, Paola and Cadamuro, Davide and Goodsell, Mark and Jaeckel, Joerg and Redondo, Javier and Ringwald, Andreas",
    title = "{WISPy Cold Dark Matter}",
    eprint = "1201.5902",
    archivePrefix = "arXiv",
    primaryClass = "hep-ph",
    journal = "JCAP",
    volume = "06",
    pages = "013",
    doi = "10.1088/1475-7516/2012/06/013",
    year = "2012"
}

@article{Irsic:2017,
    author = "Ir{\v{s}}i{\v{c}}, Vid and others",
    title = "{First Constraints on Fuzzy Dark Matter from Lyman-$\alpha$ Forest Data and Hydrodynamical Simulations}",
    eprint = "1703.04683",
    archivePrefix = "arXiv",
    primaryClass = "astro-ph.CO",
    journal = "Phys. Rev. Lett.",
    volume = "119",
    pages = "031302",
    doi = "10.1103/PhysRevLett.119.031302",
    year = "2017"
}

@article{Guo:2023,
    author = "Guo, Jinhui and He, Yuxuan and Liu, Jia and Wang, Xiao-Ping and Xie, Ke-Pan",
    title = "{Unveiling time-varying signals of ultralight bosonic dark matter at collider and beam dump experiments}",
    eprint = "2206.14221",
    archivePrefix = "arXiv",
    primaryClass = "hep-ph",
    doi = "10.1038/s42005-023-01350-6",
    journal = "Commun. Phys.",
    volume = "6",
    pages = "225",
    year = "2023"
}

@misc{Bauer:2026dgu,
    author = "Bauer, Martin and Chakraborti, Sreemanti",
    title = "{Oscillating Resonances: Imprints of ultralight dark matter at colliders}",
    eprint = "2601.19844",
    archivePrefix = "arXiv",
    primaryClass = "hep-ph",
    reportNumber = "IPPP/26/08",
    month = "1",
    year = "2026"
}

@article{Berge:2018,
    author = "Berg{\'e}, Joel and Brax, Philippe and M{\'e}tris, Gilles and Pernot-Borr{\`a}s, Martin and Touboul, Pierre and Uzan, Jean-Philippe",
    title = "{MICROSCOPE Mission: First Constraints on the Violation of the Weak Equivalence Principle by a Light Scalar Dilaton}",
    eprint = "1712.00483",
    archivePrefix = "arXiv",
    primaryClass = "gr-qc",
    doi = "10.1103/PhysRevLett.120.141101",
    journal = "Phys. Rev. Lett.",
    volume = "120",
    pages = "141101",
    year = "2018"
}

@article{Bigaran:2026,
    author = "Bigaran, Innes and Fox, Patrick J. and Gouttenoire, Yann and Harnik, Roni and Krnjaic, Gordan and Menzo, Tony and Zupan, Jure",
    title = "{Proposal for Direct Detection of Ultralight Dark Matter via Charged Lepton Flavor Violation}",
    eprint = "2503.07722",
    archivePrefix = "arXiv",
    primaryClass = "hep-ph",
    doi = "10.1103/gvf4-d2b9",
    journal = "Phys. Rev. Lett.",
    volume = "136",
    pages = "251801",
    year = "2026"
}

@article{Kobayashi:2017,
    author = "Kobayashi, Takeshi and Murgia, Riccardo and De Simone, Andrea and Ir{\v{s}}i{\v{c}}, Vid and Viel, Matteo",
    title = "{Lyman-$\alpha$ Constraints on Ultralight Scalar Dark Matter: Implications for the Early and Late Universe}",
    eprint = "1708.00015",
    archivePrefix = "arXiv",
    primaryClass = "astro-ph.CO",
    journal = "Phys. Rev. D",
    volume = "96",
    pages = "123514",
    doi = "10.1103/PhysRevD.96.123514",
    year = "2017"
}

@article{Graham:2013gfa,
    author = "Graham, Peter W. and Rajendran, Surjeet",
    title = "{New Observables for Direct Detection of Axion Dark Matter}",
    eprint = "1306.6088",
    archivePrefix = "arXiv",
    primaryClass = "hep-ph",
    doi = "10.1103/PhysRevD.88.035023",
    journal = "Phys. Rev. D",
    volume = "88",
    pages = "035023",
    year = "2013"
}

@article{Han:2012rb,
    author = "Han, Tao and Liu, Zhen",
    title = "{Potential precision of a direct measurement of the Higgs boson total width at a muon collider}",
    eprint = "1210.7803",
    archivePrefix = "arXiv",
    primaryClass = "hep-ph",
    reportNumber = "PITT-PACC-1210",
    doi = "10.1103/PhysRevD.87.033007",
    journal = "Phys. Rev. D",
    volume = "87",
    number = "3",
    pages = "033007",
    year = "2013"
}

@article{Hees:2018,
    author = "Hees, Aur{\'e}lien and Minazzoli, Olivier and Savalle, Etienne and Stadnik, Yevgeny V. and Wolf, Peter",
    title = "{Violation of the equivalence principle from light scalar dark matter}",
    eprint = "1807.04512",
    archivePrefix = "arXiv",
    primaryClass = "gr-qc",
    journal = "Phys. Rev. D",
    volume = "98",
    pages = "064051",
    doi = "10.1103/PhysRevD.98.064051",
    year = "2018"
}

@article{Touboul:2022,
    author = "Touboul, Pierre and others",
    collaboration = "MICROSCOPE",
    title = "{MICROSCOPE Mission: Final Results of the Test of the Equivalence Principle}",
    eprint = "2209.15487",
    archivePrefix = "arXiv",
    primaryClass = "gr-qc",
    journal = "Phys. Rev. Lett.",
    volume = "129",
    pages = "121102",
    doi = "10.1103/PhysRevLett.129.121102",
    year = "2022"
}

@article{Dine:2024perez,
    author = "Dine, Michael and Perez, Gilad and
              Ratzinger, Wolfram and Savoray, Inbar",
    title = "{Nelson-Barr ultralight dark matter}",
    eprint = "2405.06744",
    archivePrefix = "arXiv",
    primaryClass = "hep-ph",
    doi = "10.1103/PhysRevD.111.015049",
    journal = "Phys. Rev. D",
    volume = "111",
    number = "1",
    pages = "015049",
    year = "2025"
}

@article{Bernreuther:1991xe,
    author = "Bernreuther, W. and Botz, G. W. and Nachtmann, O. and Overmann, P.",
    title = "{CP violating effects in Z decays to tau leptons}",
    reportNumber = "HD-THEP-91-7, KA-THEP-91-1",
    doi = "10.1007/BF01562330",
    journal = "Z. Phys. C",
    volume = "52",
    pages = "567--574",
    year = "1991"
}

@article{Bauer:2024hfv,
    author = "Bauer, Martin and Chakraborti, Sreemanti and Rostagni, Guillaume",
    title = "{Axion bounds from quantum technology}",
    eprint = "2408.06412",
    archivePrefix = "arXiv",
    primaryClass = "hep-ph",
    reportNumber = "IPPP/24/53, IPPP/24/53",
    doi = "10.1007/JHEP05(2025)023",
    journal = "JHEP",
    volume = "05",
    pages = "023",
    year = "2025"
}

@article{KuraevFadin1985,
  author  = {Kuraev, E. A. and Fadin, V. S.},
  title   = {On Radiative Corrections to e+e-
             Single Photon Annihilation at High Energy},
  journal = {Sov. J. Nucl. Phys.},
  volume  = {41},
  pages   = {466--472},
  year    = {1985}
}

@misc{Du:2026ynu,
    author = "Du, Xin-Yu and Zou, Zi-Yue and He, Xiao-Gang and Xie, Keping",
    title = "{Transverse Tau Spin Correlations and the Weak Electric Dipole Moment at Future $Z$ Factories}",
    eprint = "2609.18892",
    archivePrefix = "arXiv",
    primaryClass = "hep-ph",
    month = "9",
    year = "2026"
}

@article{Belle:2021ybo,
    author = "Inami, K. and others",
    collaboration = "Belle",
    title = "{An improved search for the electric dipole moment of the $\tau$ lepton}",
    eprint = "2108.11543",
    archivePrefix = "arXiv",
    primaryClass = "hep-ex",
    reportNumber = "Belle Preprint 2021-22, KEK Preprint 2021-26",
    doi = "10.1007/JHEP04(2022)110",
    journal = "JHEP",
    volume = "04",
    pages = "110",
    year = "2022"
}

@article{Jenkins:2017dyc,
    author = "Jenkins, Elizabeth E. and Manohar, Aneesh V. and Stoffer, Peter",
    title = "{Low-Energy Effective Field Theory below the Electroweak Scale: Anomalous Dimensions}",
    eprint = "1711.05270",
    archivePrefix = "arXiv",
    primaryClass = "hep-ph",
    doi = "10.1007/JHEP01(2018)084",
    journal = "JHEP",
    volume = "01",
    pages = "084",
    year = "2018",
    note = "[Erratum: JHEP 12, 042 (2023)]"
}

@article{ALEPH:2002kbp,
    author = "Heister, A. and others",
    collaboration = "ALEPH",
    title = "{Search for anomalous weak dipole moments of the tau lepton}",
    eprint = "hep-ex/0209066",
    archivePrefix = "arXiv",
    reportNumber = "CERN-EP-2002-068",
    doi = "10.1140/epjc/s2003-01286-1",
    journal = "Eur. Phys. J. C",
    volume = "30",
    pages = "291--304",
    year = "2003"
}

@misc{Li:2025kjy,
    author = "Li, Ang and Eysermans, Jan and Bernardi, Gregorio and Dewyspelaere, Kevin and Selvaggi, Michele and Paus, Christoph",
    title = "{Model-independent ZH production cross section at FCC-ee}",
    eprint = "2512.21290",
    archivePrefix = "arXiv",
    primaryClass = "hep-ex",
    month = "12",
    year = "2025"
}

@article{Zechmeister:2009js,
    author = "Zechmeister, M. and Kurster, M.",
    title = "{The generalised Lomb-Scargle periodogram. A new formalism for the floating-mean and Keplerian periodograms}",
    eprint = "0901.2573",
    archivePrefix = "arXiv",
    primaryClass = "astro-ph.IM",
    doi = "10.1051/0004-6361:200811296",
    journal = "Astron. Astrophys.",
    volume = "496",
    pages = "577",
    year = "2009"
}

@article{Pomme:2020ipk,
    author = "Pomm{\'e}, S. and De Hauwere, T.",
    title = "{On the significance of modulations in time series}",
    doi = "10.1016/j.nima.2019.163377",
    journal = "Nucl. Instrum. Meth. A",
    volume = "956",
    pages = "163377",
    year = "2020"
}

@article{Brade:2026ncf,
    author = "Brade, Marlon P. and Juevesano, Jeremiah D. and Carbonel, Venus Abbegaile S. and Bustamante, Karen E. and Arogancia, Dennis C. and Maratas, V., Jan Mickelle",
    title = "{Search for a time-dependent Z' resonance in the dimuon channel}",
    eprint = "2606.17569",
    archivePrefix = "arXiv",
    primaryClass = "hep-ph",
    doi = "10.1103/w7cw-b316",
    journal = "Phys. Rev. D",
    volume = "114",
    number = "7",
    pages = "075033",
    year = "2026"
}

@article{Guo:2022vxr,
    author = "Guo, Jinhui and He, Yuxuan and Liu, Jia and Wang, Xiao-Ping and Xie, Ke-Pan",
    title = "{Unveiling time-varying signals of ultralight bosonic dark matter at collider and beam dump experiments}",
    eprint = "2206.14221",
    archivePrefix = "arXiv",
    primaryClass = "hep-ph",
    doi = "10.1038/s42005-023-01350-6",
    journal = "Commun. Phys.",
    volume = "6",
    number = "1",
    pages = "225",
    year = "2023"
}

@misc{Fieg:2026zdr,
    author = "Fieg, Max H. and Fox, Patrick J. and Zhang, Jinbo and Ghosh, Aishik and Varada, Virat and Whiteson, Daniel",
    title = "{Time-dependent signals of new physics at the LHC}",
    eprint = "2605.11071",
    archivePrefix = "arXiv",
    primaryClass = "hep-ph",
    reportNumber = "FERMILAB-PUB-26-0234-T",
    month = "5",
    year = "2026"
}

@article{BACON:2020ubh,
    author = "Beloy, Kyle and others",
    collaboration = "BACON",
    title = "{Frequency ratio measurements at 18-digit accuracy using an optical clock network}",
    eprint = "2005.14694",
    archivePrefix = "arXiv",
    primaryClass = "physics.atom-ph",
    doi = "10.1038/s41586-021-03253-4",
    journal = "Nature",
    volume = "591",
    number = "7851",
    pages = "564--569",
    year = "2021"
}

@article{Hees:2016gop,
    author = "Hees, A. and Gu{\'e}na, J. and Abgrall, M. and Bize, S. and Wolf, P.",
    title = "{Searching for an oscillating massive scalar field as a dark matter candidate using atomic hyperfine frequency comparisons}",
    eprint = "1604.08514",
    archivePrefix = "arXiv",
    primaryClass = "gr-qc",
    doi = "10.1103/PhysRevLett.117.061301",
    journal = "Phys. Rev. Lett.",
    volume = "117",
    number = "6",
    pages = "061301",
    year = "2016"
}

@article{Kennedy:2020bac,
    author = "Kennedy, Colin J. and Oelker, Eric and Robinson, John M. and Bothwell, Tobias and Kedar, Dhruv and Milner, William R. and Marti, G. Edward and Derevianko, Andrei and Ye, Jun",
    title = "{Precision Metrology Meets Cosmology: Improved Constraints on Ultralight Dark Matter from Atom-Cavity Frequency Comparisons}",
    eprint = "2008.08773",
    archivePrefix = "arXiv",
    primaryClass = "physics.atom-ph",
    doi = "10.1103/PhysRevLett.125.201302",
    journal = "Phys. Rev. Lett.",
    volume = "125",
    number = "20",
    pages = "201302",
    year = "2020"
}

@article{L3:1993zvy,
    author = "Adriani, O. and others",
    collaboration = "L3",
    title = "{Search for narrow vector resonances in the $Z$ mass range}",
    reportNumber = "CERN-PPE-93-37",
    doi = "10.1016/0370-2693(93)91229-G",
    journal = "Phys. Lett. B",
    volume = "313",
    pages = "326--332",
    year = "1993"
}

@article{DELPHI:1994bya,
    author = "Abreu, P. and others",
    collaboration = "DELPHI",
    title = "{Measurements of the line shape of the $Z^0$ and determination of electroweak parameters from its hadronic and leptonic decays}",
    reportNumber = "CERN-PPE-94-08, CERN-PPE-94-008, CERN-PPE-94-8",
    doi = "10.1016/0550-3213(94)90537-1",
    journal = "Nucl. Phys. B",
    volume = "417",
    pages = "3--57",
    year = "1994",
    note = "[Erratum: Nucl.Phys.B 426, 244 (1994)]"
}

@article{FCC:2025lpp,
    author = "Benedikt, M. and others",
    collaboration = "FCC",
    title = "{Future Circular Collider Feasibility Study Report: Volume 1, Physics, Experiments, Detectors}",
    eprint = "2505.00272",
    archivePrefix = "arXiv",
    primaryClass = "hep-ex",
    reportNumber = "CERN-FCC-PHYS-2025-0002",
    doi = "10.1140/epjc/s10052-025-15077-x",
    journal = "Eur. Phys. J. C",
    volume = "85",
    number = "12",
    pages = "1468",
    year = "2025",
    note = "[Erratum: Eur.Phys.J.C 86, 844 (2026)]"
}

\end{document}